\documentclass[graybox,vecphys,natbib]{svmult}

\usepackage{mathptmx}       
\usepackage{amsmath,amssymb}
\usepackage{helvet}         
\usepackage{courier}        
\usepackage{type1cm}      
\usepackage{makeidx}         
\usepackage{graphicx}        
\usepackage{multicol}        
\usepackage[bottom]{footmisc}

\makeindex     
\def \lesssim {\mathrel{\vcenter
     {\offinterlineskip \hbox{$<$}\hbox{$\sim$}}}}
\def \gtrsim {\mathrel{\vcenter
     {\offinterlineskip \hbox{$>$}\hbox{$\sim$}}}}

\begin{document}
\bibliographystyle{apa}

\title*{High Energy Cosmic Rays From Supernovae}
\label{Chapter:CosmicRaysfromSupernovae}
\author{Giovanni Morlino}
\institute{Giovanni Morlino \at Gran Sasso Science Institute, 
Viale F. Crispi~7, I-67100 L'Aquila, Italy\\ \email{giovanni.morlino@gssi.infn.it}
}
%
%
\maketitle

\abstract*{[SAME AS ABSTRACT BELOW]
}

\abstract{Cosmic rays are charged relativistic particles that reach the Earth with extremely high energies, providing striking evidence of the existence of effective accelerators in the Universe. Below an energy around $\sim10^{17}$ eV  cosmic rays are believed to be produced in the Milky Way while above that energy their origin is probably extragalactic. In the early '30s supernovae were already identified as possible sources for the Galactic  component of cosmic rays.  After the '70s this idea has gained more and more credibility thanks to the the development of the {\it diffusive shock acceleration} theory, which provides a robust theoretical framework for particle energization in astrophysical environments. Afterwards, mostly in recent years, much observational evidence has been gathered in support of this framework, converting a speculative idea in a real paradigm.
In this Chapter the basic pillars of this paradigm will be illustrated. This includes the acceleration mechanism, the non linear effects produced by accelerated particles onto the shock dynamics needed to reach the highest energies, the escape process from the sources and the transportation of cosmic rays through the Galaxy. The theoretical picture will be corroborated by discussing several observations which support the idea that supernova remnants are effective cosmic ray factories.
}

\section{Introduction}
\label{sec:morlino-intro}

\begin{figure}
\begin{center}
\includegraphics[scale=.5]{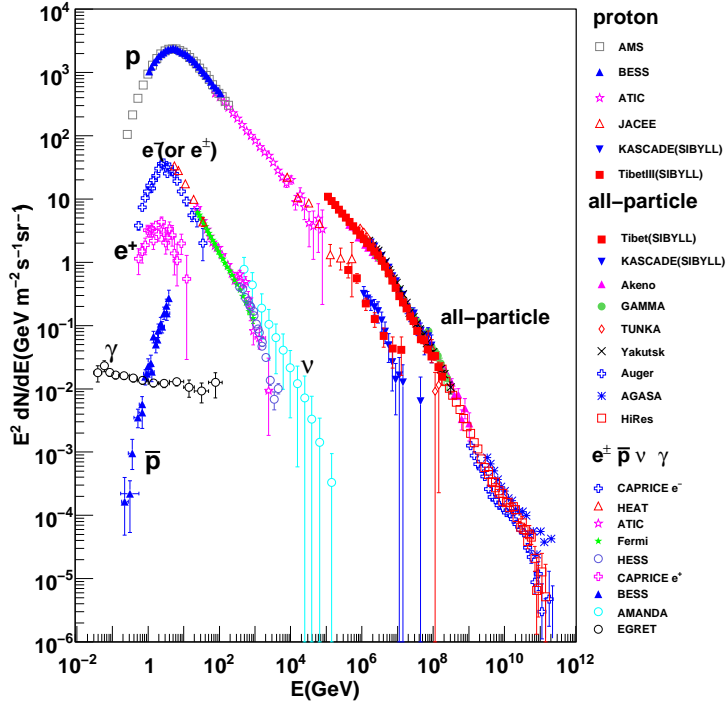}
\end{center}
\caption{Spectrum of cosmic rays at the Earth. The all-particle spectrum measured by different experiments is plotted, together with the proton spectrum. The subdominant contributions from electrons, positrons and antiprotons as measured by the PAMELA experiment are shown. For comparison also atmospheric neutrino and diffuse gamma-ray background are shown.}
\label{fig:CRspectrum}
\end{figure}

Cosmic Rays (CR) are charged particles detected at the Earth, or in the space just around the Earth, mainly consisting of protons (hydrogen nuclei) with about $10\%$ fraction of helium nuclei and smaller abundances of heavier elements.
The flux of all nuclear components is shown in Fig.\ref{fig:CRspectrum}.  
In spite of the fact that the CR spectrum extends over at least 13 decades in energy, extracting information from it is hard because it is nearly featureless. For energies greater than $\sim 30$~GeV, where the Solar wind screening effect becomes negligible, the spectrum resembles to a broken power-law with a spectrum changing from $\propto E^{-2.7}$ to $\propto E^{-3.1}$ at an energy of $E_{\rm knee} \approx 3\times 10^{15}$~eV (a feature called the  \textit{knee}. 
A second change in the spectrum occurs around $E_{\rm ankle} \approx 3 \times 10^{18}$~eV where the slope flattens again towards a value close to $2.7$ (usually referred to as the \textit{ankle}).  In the highest energy region the flux falls to very low values, hence measurement becomes extremely difficult (at $3\times10^{20}$~eV the flux is 1 particle per km$^{2}$ each 350 years).

There is evidence that the chemical composition of CRs changes across the knee region with a trend to become increasingly more dominated by heavy nuclei at high energy \citep[see][for a review]{Hoerandel:2006}, at least up to  $\sim 10^{17}$ eV. This evidence could be explained if the CR acceleration mechanism were rigidity-dependent and  the maximum energy of protons could reach $3\times 10^{15}$ eV. Then heavier nuclei, with charge $Z$, would reach $Z$ times larger energies. In this scheme the heaviest nuclei, namely Fe, have an energy of $26 \times E_{\rm knee}$ and the knee structure results as the superposition of the cut-offs of different species.

CRs up to an energy around $10^{17}$~eV are believed to originate in our own Galaxy. On the contrary, particles with energy beyond the ankle, usually referred to as \textit{Ultra-High Energy Cosmic Rays} (UHECRs), cannot be confined in the Galaxy, because their Larmor radius in the typical Galactic magnetic field is of the same order of the Galaxy size, or even larger. Hence, if they were produced in the Galaxy, the particle deflection would be small enough that the arrival direction should trace the source's position in the sky. On the contrary, the incoming spatial distribution of UHECRs is nearly isotropic, hence the general opinion is that these particles come from extragalactic sources.

A connection between CRs and supernovae (SN) was firstly proposed in the early '30s \citep{BaadeZwicky:1934} on the basis of a simple energetic argument. The power needed to maintain the Galactic CRs at the observed level against losses due to escape from the Galaxy can be estimated as follows
\begin{equation}
  P_{\rm CR}\sim U_{\rm CR} V_{\rm CR}/\tau_{\rm res} \approx 10^{40} {\rm erg/s}\,,
\end{equation}
where $U_{\rm CR}\approx 0.5$~eV/cm$^3$ is the CRs energy density measured at the Earth and $V_{\rm CR}\sim 400$~kpc$^3$ is the volume of the Galactic halo where CRs are efficiently confined. The typical residence time of a cosmic ray in the Galaxy we assume to be $\tau_{\rm esc}\sim 5\times 10^6$ yr (see Section~\ref{sec:morlino-propagation}).
Now we know that the energy released by a single supernova explosion in kinetic energy of the expanding shell is around $10^{51}$ erg, therefore the total energy injected into the Galactic environment is 
\begin{equation}
  P_{\rm SNR}= R_{\rm SN} E_{\rm SNR} \approx 3 \times 10^{41} {\rm erg/s} \,.
\end{equation}
where $R_{\rm SN}\sim 0.03$~yr$^{-1}$ is the rate of supernova explosion in the Galaxy. Accounting also for the uncertainties in the parameters, the energy density of the Galactic CRs component can be explained if one assumes that a fraction around $3-30\%$ of the total supernovae mechanical energy is transferred to non-thermal particles.

This energetic argument was the only basis in favor of the SN hypothesis until the '70s, when a mechanism able to transfer energy from  SNe to non-thermal particles was proposed, namely the stochastic acceleration occurring at the SNR shocks.
Since then, this idea have received more and more attention thanks to a number of observations, especially in radio, X-rays and gamma-rays, that confirmed many predictions and triggered further improvements of the theory. 

In this Chapter we summarize the basic theoretical aspects of the SNR-CR connection. Section~\ref{sec:morlino-DSA} is devoted to explain the diffusive shock acceleration in the test-particle limit while in Section~\ref{sec:morlino-NLDSA} we discuss how non linear effects produced by accelerated particles can modify the shock structure. In Sections~\ref{sec:morlino-escaping} and \ref{sec:morlino-propagation} we discuss the escaping process from the sources and the diffusion through the Galaxy, respectively. Finally in Section~\ref{sec:morlino-obs} a number of relevant observations are discussed. Conclusions and future prospectives are drawn in Section~\ref{sec:morlino-conclusions}.

\section{The acceleration mechanism}
\label{sec:morlino-DSA}

\subsection{First and second order Fermi acceleration}
\label{sec:morlino-Fermi}
\begin{figure}[t]
\includegraphics[scale=.21]{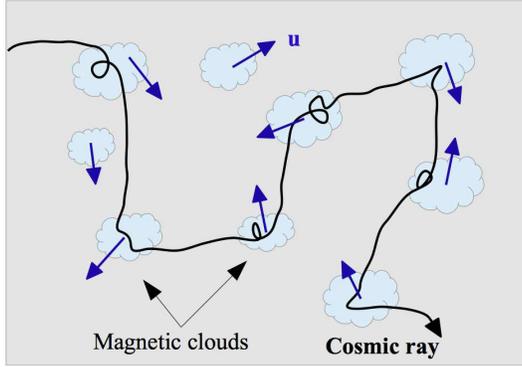}
\sidecaption[t]
\caption{Schematic representation of the original Fermi idea to energize cosmic rays through repeated scatters with magnetic clouds randomly moving in the Galaxy.}
\label{fig:Fermi-II}
\end{figure}

The most common invoked acceleration mechanism in astrophysics is diffusive shock acceleration (DSA) also called the $1^{st}$ order Fermi process. In fact the seminal idea was put forward by \cite{Fermi:1949,Fermi:1954} who proposed that CRs could be accelerated by repeated stochastic scattering in a turbulent magnetic field which Fermi idealized as magnetized clouds moving around the Galaxy with random velocity (see Fig.~\ref{fig:Fermi-II}). In the form in which Fermi first put it forward, this idea, today called $2^{nd}$ order Fermi process, does not work either to explain the shape of the CRs spectrum, or to account for their total energy density. Nevertheless, it well illustrates the concept of stochastic acceleration, hence it is worth to be discussed here.

Let us consider a single particle with energy $E_1$, in the Galaxy's frame, and a cloud with Lorentz factor $\gamma$ and speed $u=\beta c$. For simplicity we assume that the particle is already relativistic, i.e. $E\sim pc$. In the reference frame of the cloud the energy is
\begin{equation}
 E_1'= \gamma E_1 (1-\beta \cos \theta_1) \,,
\end{equation}
where $\theta_1$ is the angle between particle's and cloud's velocities. After the interaction the energy in the cloud's frame remains unchanged, namely $E_2'=E_1'$, while the final energy in the Galaxy's frame is 
\begin{equation}
 E_2= \gamma E_2' (1+\beta \cos \theta_2') \,,
\end{equation}
where $\theta_2'$ is the exit angle in the cloud's frame. Hence after a single encounter the energy gain is
\begin{equation} \label{eq:gain}
 \frac{\Delta E}{E_1} \equiv \frac{E_2-E_1}{E_1}  =
 \frac{1-\beta \cos\theta_1 + \beta \cos\theta_2'-\beta^2 \cos\theta_1 \cos\theta_2'}{1-\beta^2} - 1 \,.
\end{equation}
To get the mean energy gain we need to average over the incoming and the outcoming directions. Because the scattering in the cloud frame is isotropic, we have $\langle \cos\theta_2'\rangle=0$. On the other hand, the mean incoming direction can be computed averaging over the particle flux, which is proportional to the relative velocity, $\beta_r= 1-\beta \cos\theta_1$. Hence, if the particle distribution is isotropic in the Galaxy's frame, we simply have
\begin{equation}
 \langle \cos\theta_1\rangle = \frac{\int d\Omega \,\beta_r
   \cos\theta_1}{\int d\Omega \, \beta_r} 
 = \frac{\int_{-1}^{1} d\cos\theta_1 \,(1-\beta \cos\theta_1)
   \cos\theta_1}{\int_{-1}^{1} d\cos\theta_1 \,(1-\beta \cos\theta_1)} 
 = -\frac{\beta}{3}
\end{equation}
and the average energy gain become
\begin{equation}
 \frac{\Delta E}{E_1} = \frac{1+\frac{1}{3}\beta^2}{1-\beta^2} -1 
 \simeq \frac{4}{3} \beta^2 \,.
\end{equation}
The last passage is obtained assuming that $\beta \ll 1$. In spite of the fact that in each interaction a particle can either gain or lose energy, the average energy gain is positive simply because the cloud is moving, hence the flux of particle crossing the cloud in front is greater then the one leaving the cloud from behind. The proportionality of the energy gain to the second power of the speed justifies the name of $2^{nd}$ order Fermi mechanism and this is exactly the reason why it cannot explain the CR spectrum. In fact the random velocities of clouds are relatively small, $v/c\approx 10^{-4}$ and, for a particle with a mean free path of 0.1 pc, the collisions would like to occur only few times per year hence the final energy gain is really modest. Moreover the predicted energy spectrum strongly depends on the model details \cite[see][Ch.~7]{Longair:1992} a conclusion at odds with observations.

As recently as the seventies, several authors independently realized that when Fermi's idea is applied to particles in the vicinity of a shock wave, the result changes dramatically \citep{Skilling:1975a,Skilling:1975b,Axford:1977,Krymskii:1977,Bell:1978a,Bell:1978b,Blandford-Ostriker:1978}. This time the magnetic turbulence in the plasma provides the scattering centers needed to confine particles around the shock wave, allowing them to cross the shock repeatedly. Each time a particle crosses the shock front, it always suffers head-on collisions with the magnetic turbulence on the other side of the shock, gaining a bit of energy which is subtracted from the bulk motion of the plasma. Let us describe this process with more details.

\begin{figure}[t]
\begin{center}
\includegraphics[width=0.99\linewidth]{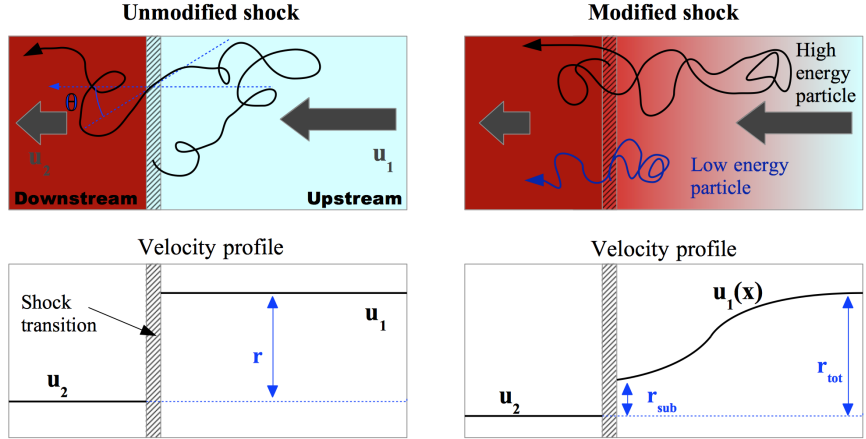}
\end{center}
\caption{{\bf Left}. Structure of an unmodified plane shock wave. Particle diffusing from upstream towards downstream feel the compression factor $r$ in the velocity of the plasma, which is the same at all energies.
{\bf Right}. Shock structure modified by the presence of accelerated particles. The pressure exerted by accelerated particles diffusing upstream slows down the plasma creating a ``precursor''. High energy particles, which propagate farther away from the shock, feel now a larger compression factor with respect to low energy particles which diffuse closer to the shock.}
\label{fig:shock_structure}
\end{figure}

Consider a plane shock moving with velocity $u_{\rm sh}$. In the frame where the shock is at rest the upstream plasma moves towards the shock with velocity $u_1 \equiv u_{\rm sh}$, while the downstream plasma moves away form the shock with velocity $u_2$ (see Fig.~	\ref{fig:shock_structure}, left panel). The situation is similar to what happens in the case of a moving cloud described before, but this time the relative velocity between downstream and upstream plasma is $u_r \equiv \beta_r c = u_1-u_2$. Assuming that the particles density, $n$, is isotropic, the flux of particle crossing the shock from downstream region towards the upstream one is 
\begin{equation} \label{eq:flux_Jmin}
 J_-= \int \frac{d\Omega}{4\pi} \, n c \cos\theta = \frac{nc}{4} \,,
\end{equation} 
where the integration is performed in the interval $-1 \leqslant \cos\theta \leqslant 0$. Hence the average value of the incoming angle is 
\begin{equation} \label{eq:cosTheta1}
 \langle \cos\theta_1 \rangle = \frac{1}{J_-}\int \frac{d\Omega}{4\pi} \, n
 c \cos^2\theta_1 = -\frac{2}{3} \,
\end{equation} 
while for the outcoming direction we have $\langle \cos\theta_2' \rangle= 2/3$ because the integration is performed for $0 \leqslant  \cos\theta \leqslant 1$. According to Eq.~(\ref{eq:gain}) the average energy gain in a single cycle downstream-upstream-downstream is:
\begin{equation} \label{eq:E_gain}
 \frac{\Delta E}{E} =
 \frac{1+\frac{4}{3}\beta_r+\frac{4}{9}\beta_r^2}{1-\beta_r^2} -1 
 \sim \frac{4}{3} \, \beta_r \,.
\end{equation}
Compared to the collision with clouds, the shock acceleration is more efficient, resulting in an energy gain proportional to the relative velocity between upstream and downstream plasmas, hence the name \textit{first order} Fermi process. 


\subsection{Particle spectrum}
\label{sec:morlino-DSAspectrum}
The most remarkable property of the first order Fermi mechanism consists in the production of a particle spectrum which is a universal power law. Such universality is a consequence of the balance between the energy gain and the escape probability from the accelerator, as we illustrate below. 

During each cycle around the shock, a particle has a finite probability to escape because of the advection with the downstream plasma. In a steady state situation the particle flux advected towards downstream infinity is simply $J_\infty= n u_2$, while no particle can escape towards upstream infinity (in reality this assumption can be violated for particles at maximum energy, see Section~\ref{sec:morlino-escaping}). For the flux conservation we have $J_+= J_- + J_\infty$, where $J_+$ and $J_-$ are the flux of particles crossing the shock from upstream towards downstream and vice versa. Using Eq.~(\ref{eq:flux_Jmin}) the escape probability can be expressed as
\begin{equation} \label{eq:Pesc}
 P_{\rm esc}= \frac{J_{\infty}}{J_+}= \frac{J_{\infty}}{J_{\infty} + J_{-}} \simeq \frac{4u_2}{c} \,
\end{equation}
and is independent from the particle's energy.

Now, to calculate the particle spectrum, we can use the microscopic approach, following the fate of a single particle which enters the acceleration process. For each cycle the energy gain, $\xi \equiv \Delta E/E$, is given by Eq.~(\ref{eq:E_gain}) and it is independent of the initial energy, $E_0$. After $k$ cycles the particle's energy will be $E= E_0(1+\xi)^k$, implying that the number of cycles needed to reach an energy $E$ is equal to 
\begin{equation} \label{eq:n}
 k= \frac{\ln(E/E_0)}{\ln(1+\xi)} \,.
\end{equation}
Moreover after each cycle the particle has a probability $1-P_{\rm esc}$ to undergo another acceleration cycle. Hence, after $k$ cycles the number of particles with energy greater then $E$ is proportional to
\begin{equation} \label{eq:N>E}
 N(>E) \propto \sum_{i=k}^{\infty} (1-P_{\rm esc})^i
 	= \frac{(1-P_{\rm esc})^k}{P_ {\rm esc}}
   	= \frac{1}{P_ {\rm esc}} \left(\frac{E}{E_0}\right)^{-\delta} \,,
\end{equation}
where $\delta = - \frac{\ln(1-P_{\rm esc})}{\ln(1+\xi)}$. Because both $P_{\rm esc}$ and $\xi$ are small quantities, we can approximate $\delta \simeq P_{\rm esc}/\xi$. Deriving Eq.~(\ref{eq:N>E}) we get the differential energy spectrum which is a simple power law, $f(E)= dN/dE \propto E^{-\alpha}$ where the spectral index can be calculated using Eqs.~(\ref{eq:E_gain}) and (\ref{eq:Pesc}):
\begin{equation} \label{eq:slopeDSA}
 \alpha = 1+\delta \simeq 1+P_{\rm esc}/\xi   =  1 + \frac{3 u_2}{u_1-u_2} = \frac{r+2}{r-1} \,. 
\end{equation}
The last equality makes use of the compression ratio, $r \equiv u_1/u_2$, that  can be obtained using the flux conservation of mass, momentum and energy across the shock discontinuity. For a non relativistic hydrodynamical shock propagating with a  Mach number $M=u_{\rm sh}/v_{\rm sound}$ into a gas with adiabatic index $\gamma_g$, the very well known result is:
\begin{equation} \label{eq:r}
 r = \frac{(\gamma_g+1)M^2}{(\gamma_g-1)M^2+2} \,.
\end{equation}
For very strong shocks ($M\gg 1$) and an ideal monoatomic gas ($\gamma_g=5/3$) the compression factor reduces to $4$ and  the predicted particle spectrum becomes
\begin{equation} \label{eq:DSAspectrum}
  f(E) \propto E^{-2} \,.
\end{equation}
The particle spectrum is often expressed in momentum rather than energy, the relation between the two being $f(E) dE = 4 \pi f(p) p^2dp$. Hence $f(E)\propto E^{-2}$ means $f(p) \propto p^{-4}$. Such universal spectrum is based on two ingredients: 1) the energy gained in a single acceleration cycle is proportional to the particle's energy and 2) the escaping probability is energy independent. Both these properties are direct consequences of the underlying assumption that the particle transport is diffusive, even if the details of the scattering process never enter the calculation. For this reason the $1^{st}$ order Fermi mechanism is also called {\it diffusive shock acceleration} (DSA). From a mathematical point of view the diffusion guarantees the isotropization of the particle distribution both in the upstream and downstream reference frames. If this where not the case, Eq.~(\ref{eq:E_gain}) would not hold any more. On the other hand, from a physical point of view the scattering process between particles and magnetic turbulence is the real responsible for the energy transfer between the plasma bulk kinetic energy and the non-thermal particles.


\subsection{Particle diffusion in weak magnetic turbulence}
\label{sec:morlino-Diffusion}
\begin{figure}[t]
\includegraphics[scale=.33]{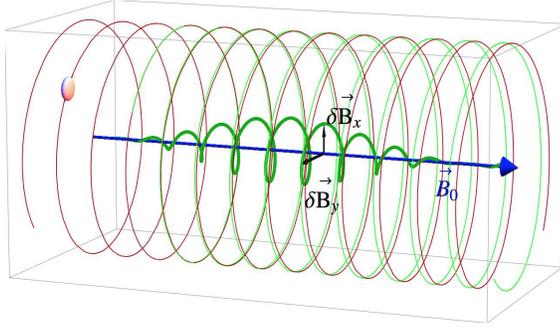}
\sidecaption[t]
\caption{The red line shows the motion of a particle in a large scale magnetic field while the green line shows the motion when a small perturbation $\delta \bf B \perp B_0$ is added on top of $\bf B_0$.}
\label{fig:particle}
\end{figure}
The assumption of diffusive motion used to derive the universal spectrum in DSA deserves a deeper discussion.
The description of charged particle motion in a plasma with generic magnetic turbulence is a very complicated task and represents an active area of research \cite[see][]{Shalchi:2009}. Here we limit our attention to an idealized situation where a particle moves in presence of a regular magnetic field $B_0$ on top of which there are small perturbations $\delta B \perp B_0$ (see Fig.~\ref{fig:particle}). In this case the motion can be easily described in the quasi-linear regime, namely when $\delta B \ll B_0$. In absence of perturbations the particle simply gyrates along $B_0$ with frequency $\Omega = q B_0/(mc \gamma)$. When a perturbation is added such that its  wave-length is of the same order of the particle Larmor radius $r_{L}=v/ \Omega$, the particle ``resonate'' with the perturbation and its pitch angle $\theta$ suffer a small deviation \cite[see][section~3.2 for a complete derivation]{Blasi_rew:2013}. 
If $P(k) dk$ is the wave energy density in the wave number range $dk$ at the resonant wave number $k=\Omega/v \cos(\theta)$, the total scattering rate can be written as:
\begin{equation} \label{eq:nu_scatter}
\nu_{\rm sc} = \frac{\pi}{4} \left( \frac{kP(k)}{B_{0}^{2}/8\pi}\right) \Omega.
\end{equation}
The time required for the particle direction to change by $\delta\theta\sim 1$ is  $\tau\sim 1/\nu_{\rm sc}$ and the mean free path needed to reverse the velocity direction along $B_0$ is $\lambda_{\rm mfp}=v\tau$, so that the spatial diffusion coefficient can be estimated as
\begin{equation} \label{eq:diff_coeff}
D(p) = \frac{1}{3} v \lambda_{\rm mfp} \simeq  \frac{1}{3} v^{2} \Omega^{-1}\left( \frac{kP(k)}{B_{0}^{2}/8\pi}\right)^{-1} = \frac{1}{3} \frac{r_{L}v}{{\cal F}},
\end{equation}
where ${\cal F}=\left( \frac{kP(k)}{B_{0}^{2}/8\pi}\right)$ is the normalized energy density per unit logarithmic bandwidth of magnetic perturbations. Notice that in general ${\cal F} \ll 1$ for the turbulence in the interstellar medium. If ${\cal F} \sim 1$ the diffusion coefficient approaches the so called Bohm limit, defined as $D_{\rm Bohm} \equiv r_L v/3$ which is usually assumed as the smallest possible diffusion coefficient. Beyond ${\cal F} \sim 1$ the turbulence becomes strongly non-linear and many different phenomena can affect the particle motion other than the resonant scattering.

\subsection{Maximum energy}
\label{sec:morlino-Emax}
We saw in the Introduction that the  knee structure of the CR spectrum requires protons to be accelerated up to $E_{p,\max} \approx 3\times 10^{15}$ eV and that the maximum energy of heavier nuclei should scales with their nuclear charge.
The maximum achievable energy depends on the balance between the acceleration time and the minimum between the energy loss time and the age of the accelerator. In the context of a SNR shock, energy losses for hadrons do not represent a strong constraint, while for electrons both synchrotron and inverse Compton process can be fast enough to limit the acceleration process. Here we focus our attention only on hadrons.

We start noticing that even if the particle spectrum predicted by the DSA is completely insensitive to the scattering properties, the acceleration time does depend on scattering in that it determines the time it takes for the particles to get back to the shock.
In the assumption of isotropy, the flux of particles that cross the shock from downstream to upstream is $n c/4$ (see Eq.~(\ref{eq:flux_Jmin})), which means that the upstream section is filled through a surface $\Sigma$ of the shock in one diffusion time upstream with a number of particles $n(c/4) \tau_{\rm diff,1} \Sigma$ ($n$ is the density of accelerated particles at the shock). This number must equal the total number of particles within a diffusion length upstream $L_1 = D_1/u_1$, namely:
\begin{equation}
  \frac{n c}{4} \Sigma \tau_{\rm diff,1} = n \Sigma \frac{D_1}{u_1}
\end{equation}
which implies for the diffusion time upstream $\tau_{\rm diff,1} = 4D_1/(c u_1)$ . A similar estimate downstream leads to $\tau_{\rm diff,2} = 4D_2/(c u_2)$, so that the duration of a full cycle across  the shock is $\tau_{\rm cycle} = \tau_{\rm diff,1} + \tau_{\rm diff,2}$. The acceleration time is now:
\begin{equation} \label{eq:t_acc}
  t_{\rm acc} = \frac{t_{\rm cycle}}{\Delta E/E} = \frac{3}{u_1-u_2} \left( \frac{D_1}{u_1} + \frac{D_2}{u_2} \right)
  	            \approx 8 \, \frac{D_1}{u_{\rm sh}^2} \,,
\end{equation}
where the last passage is obtained assuming that the upstream turbulence is compressed at the shock by the same compression factor of the plasma $\delta B_2 \approx 4 \delta B_1$.
The maximum achievable energy is then determined by the condition $t_{\rm acc}(E_{\max}) = t_{\rm SNR}$.

SNR shocks remain efficient accelerators only for a relatively short time. Immediately after the SN explosion, the SN ejecta expand in the ISM with a velocity which is almost constant and highly supersonic. During this phase, the so-called ejecta-dominated phase, acceleration is expected to be effective because the shock speed remains almost constant. After some time, however, the mass of the circumstellar medium that the shock sweeps up becomes comparable to the mass of the ejecta, and, from that point on, the remnant enter the Sedov-Taylor phase where the shock velocity starts to decrease. This happens at a time $t_{\rm ST} = R_{\rm ST}/u_{\rm sh}$, where the shock speed can be determined from the condition $(1/2)M_{\rm ej}V_{\rm ej} = E_{\rm SN}$ and the ejecta velocity is $V_{\rm ej}=u_{\rm sh}/4$, while the radius of the remnant is defined by the condition that the swept up mass is equal to the mass of the ejecta, namely $(4\pi/3) \rho_{\rm ISM} R_{\rm ST}^3 = M_{\rm ej}$. One finds:
\begin{equation} \label{eq:t_ST}
  t_{\rm ST} \approx 50 \; \left( \frac{M_{\rm ej}}{M_{\odot}} \right)^{\frac{5}{6}} \, 
  				\left( \frac{E_{\rm SN}}{10^{51} \rm erg} \right)^{-\frac{1}{2}} \, 
				\left( \frac{n_{\rm ISM}}{\rm cm^{-3}} \right)^{-\frac{1}{3}} \, {\rm yr},
\end{equation}
For typical values of the parameters, the Sedov-Taylor phase starts after only 50-200 years. 
Now, equating Eqs.~(\ref{eq:t_acc}) and (\ref{eq:t_ST}) and using the result for the diffusion coefficient from Eq.~(\ref{eq:diff_coeff}) we get an estimate for the maximum energy:
\begin{equation} \label{eq:E_max}
  E_{\max} = 5 \times 10^{13} \, Z \, {\cal F}(k_{\min}) \,
  		   \left( \frac{B_0}{\mu \rm G} \right)
                    \left( \frac{M_{\rm ej}}{M_{\odot}} \right)^{-\frac{1}{6}} 
                    \left( \frac{E_{\rm SN}}{10^{51} \rm erg} \right)^{\frac{1}{2}} 
		   \left( \frac{n_{\rm ISM}}{\rm cm^{-3}} \right)^{-\frac{1}{3}} \; {\rm eV} \,,
\end{equation}
where $k_{\min} = 1/r_L(E_{\max})$ is the wave number resonant with particles at maximum energy.
We notice that more realistic estimates of the maximum energy (for example accounting for the fact that the shock speed is sligthly decreasing also during the ejecta-dominated phase) usually return somewhat lower values. 

A few comments are in order. First of all Eq.(\ref{eq:E_max}) has the desired proportionality to the particle charge, $Z$, which is a property required to explain the knee feature. Nevertheless, the maximum energy of protons could reach $E_{\rm knee}$ only if ${\cal F}(k_{\min}) \gg 1$, namely the magnetic turbulence at the scale of $r_L(E_{\max})$ must be much larger than the pre-existing field, $\delta B \gg B_0$. Clearly if this condition were realized, the linear theory used to derive the diffusion coefficient in Section~{\ref{sec:morlino-Diffusion} would not hold anymore. 
Apart from that, the value of turbulence in the ISM at scales relevant for us is $\delta B/B_0 \lesssim 10^{-4}$ \citep{Armstrong-CR:1981}, hence, in absence of any mechanism able to amplify the magnetic turbulence, SNR shocks could accelerate protons only up to the irrelevant energy of a few GeV. 

One should keep in mind, however, that the magnetic amplification can increase $E_{\max}$ only if it occurs both upstream and downstream of the shock otherwise particles could escape either from one side or the another. Having a magnetic amplification downstream is quite an easy task, in fact the shocked plasma is usually highly turbulent and hydrodynamical instabilities can trigger the amplification, converting a fraction of the turbulent motion into magnetic energy \citep{Giacalone-Jokipii:2007}. Conversely, there are no reasons, in general, to assume that the plasma where a SNR expands is highly turbulent to start with.

This puzzle has been partially solved by the idea that the same accelerated particles can amplify the magnetic field upstream while they try to diffuse far away from the shock \citep{Skilling:1975a,Skilling:1975b,Bell:1978a,Bell:1978b,Lagage-Cesarsky:1983a,Lagage-Cesarsky:1983b}. Nevertheless this idea in its original form can only produce $\delta B \lesssim B_0$, i.e. ${\cal F} \lesssim 1$, resulting in a maximum energy for protons of $10-100$ TeV. Theoretically speaking, a big effort is needed to fill up the last decade of energy to reach $E_{\rm knee}$. The solution to this conundrum probably resides in the non linear effects of DSA, as will be illustrate in Section~\ref{sec:morlino-NLDSA}.

A final comment concerns the parameter values used in Eq.(\ref{eq:E_max}), typical for a type Ia SNe, which have $M_{\rm ej} \approx M_{\odot}$ and expand in the ISM whose typical density is $n_{\rm ISM} \approx 0.1-1$ cm$^{-3}$. Remarkably, $E_{\max}$ is only weakly dependent on those parameters, hence its value does not change significantly when one considers core collapse SNe, which have $M_{\rm ej}\approx 10 M_{\odot}$ and expand inside the diluted bubbles ($n\sim 0.01$ m$^{-3}$) inflated by the wind of the progenitor star.


\section{DSA in the non linear regime}
\label{sec:morlino-NLDSA}
The DSA illustrated in Section~\ref{sec:morlino-DSA} assumes that the amount of energy transferred from the shock to non-thermal particles is only a negligible fraction of the plasma kinetic motion. There are several arguments supporting the idea that this condition is violated in SNR shocks. When the back reaction of accelerated particles is taken into account the DSA becomes a non linear theory (NLDSA): shock and accelerated particles become a symbiotic self-organizing system and require sophisticated mathematical tools to be studied \cite[see][for a review on mathematical aspects of NLDSA]{Malkov-Drury:2001}. 
Even more interestingly, NLDSA makes many predictions which seem to be supported by observations \cite[see][for a discussion of the observational evidence of NLDSA]{Blasi_rew:2013, Amato_rew:2014}. Here we summarize the main features of NLDSA, underlining the aspects which are still under investigation.

\subsection{The dynamical reaction of accelerated particles}
\label{sec:morlino-NLDSAp}
As discussed in the Introduction, if SNRs are the main sources of Galactic CRs then a fraction $\sim 10\%$ of their kinetic energy needs to be transferred to CRs. This means that during the acceleration process, the diffusion of CRs ahead of the shock exerts a non negligible pressure onto the incoming plasma, slowing it down (in the rest frame of the shock) and creating a precursor (see right panel of Fig.~\ref{fig:shock_structure}). Indeed, the estimate of $10\%$ takes into account the entire lifetime of the remnant, hence the instantaneous efficiency could even be larger, because SNRs likely accelerate CRs efficiently only during a fraction of their life. 
The CR pressure is expressed as follows:
\begin{equation} \label{eq:Pcr}
  P_{\rm CR} = \frac{4 \pi}{	3}\int_{p_{\min}}^{p_{\max}} p^2 dp pc f_{\rm CR}(p) \,. 
\end{equation}
where $f_{\rm CR}(p)$ is the CR distribution as a function of momentum. The shock acceleration efficiency is usually defined in terms of pressure normalized to the incoming ram pressure of the plasma, namely $\xi_{\rm CR} \equiv P_{\rm CR} / (\rho u_{\rm sh}^2)$. 

Now, a correct description of the acceleration process requires that $P_{\rm CR}$ is included in the energy and momentum equations of the shock dynamics. This leads to a  compression factor which depends on the location upstream of the shock. Particles with different energies feel now a different  compression factor which increases for larger energies (compare left and right panels in Fig.\ref{fig:shock_structure}). Moreover, when the highest energy particles escape from the acceleration toward upstream infinity, the shock becomes {\it radiative} thereby inducing an increase of the total compression factor between upstream infinity and downstream. As a consequence the predicted spectrum is no more a straight power law but becomes curved, with a spectral index which changes with energy, being steeper than 2 for lower energies and harder than 2 for the highest energies (see the example in Fig.\ref{fig:modified-spectrum}).

This prediction is somewhat at odds with observations. Even if a small curvature has been inferred from the synchrotron spectrum of few young SNRs \citep{Reynolds-Ellison:1992}, NLDSA predicts total compression ratios $\gg 4$ and, consequently, spectra much harder than $\propto E^{-2}$ (see Eq.~(\ref{eq:slopeDSA})). The solution to this inconsistency is found in a second aspect of the non-linearity, namely the dynamical reaction of magnetic field. Before discussing this aspect in Section~\ref{sec:morlino-MFreaction}, we give a closer look to the process of magnetic field amplification.

\begin{figure}[t]
\includegraphics[scale=.35]{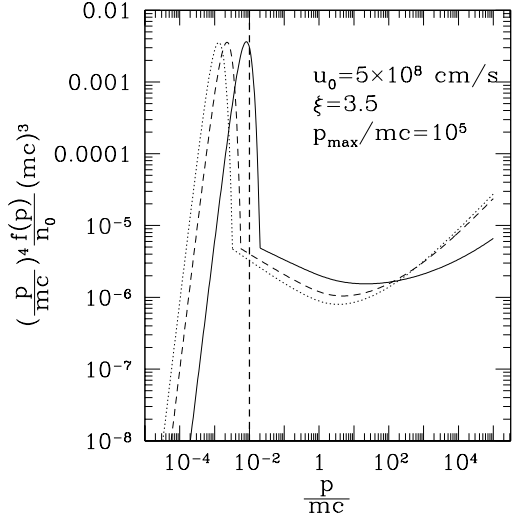}
\sidecaption[t]
\caption{Particle spectra resulting from a CR modified shock  with Mach number $M_0 = 10$ (solid line), $M_0 = 50$ (dashed line) and $M_0 = 100$ (dotted line). The vertical dashed line is the location of the thermal peak as expected for an ordinary shock with no particle acceleration. The shock velocity is $5000$ km s$^{-1}$ an the maximum energy is $c p_{\max}= 10^5$ GeV \citep{Blasi-GV:2005}.}
\label{fig:modified-spectrum}
\end{figure}

\subsection{Magnetic field amplification}
\label{sec:morlino-MFA}
Magnetic field amplification is probably the most relevant manifestation of NLDSA. Whenever energetic particles stream faster than the  Alfv\'en speed, they generate  Alfv\'en waves with wavelength close to their gyro-radius and with the same helicity of the particle motion. This instability is called  {\it resonant streaming instability} \citep{Skilling:1975b,Bell:1978a,Achterberg:1983} and can allow small perturbation to be amplified by several orders of magnitude. 

The mechanism, applied to the case of shocks, can be understood as follows. As soon as particles cross the shock discontinuity from downstream towards upstream, in the upstream rest frame they stream with a velocity $V_d = u_{\rm sh}$ and carry a total momentum $P_{\rm CR} = n_{\rm CR} m \gamma_{\rm CR} V_d$, where $\gamma_{\rm CR}$ is the CR Lorentz factor. Due to the scattering process, the distribution function is isotropized in the rest frame of the waves on a typical time scale given by the inverse of Eq.~(\ref{eq:nu_scatter}) and the total momentum reduces to $n_{\rm CR} m \gamma_{\rm CR} V_A$ ($\gamma_{\rm CR}$ does not change because magnetic field does not make work). Hence the rate of momentum loss is
\begin{equation} \label{eq:Pcr_lost}
  \frac{d P_{\rm CR}}{dt} = \frac{\Delta P_{\rm CR}}{\tau} = \frac{n_{CR} m \gamma_{\rm CR}}{\tau} \left( V_d -V_A \right) \,. 
\end{equation}
Such momentum is transferred to Alfv\'en waves which move at speed $V_A=B_0/\sqrt{4\pi \rho_i}$ where $\rho_i$ is the density of thermal ions. The transport equation for magnetic pressure in presence of amplification can be written as
\begin{equation} \label{eq:Pw_gain}
  V_A \frac{d P_{w}}{dt} = \Gamma_{\rm res} \frac{\delta B^2}{8\pi} \,,
\end{equation}
Assuming equilibrium between the momentum lost by CRs and the momentum gained by the waves, we get the growth rate for the waves: 
\begin{equation} \label{eq:Gamma_res}
  \Gamma_{\rm res} = \frac{\pi}{2} \Omega_{\rm ci} \frac{n_{\rm CR}(p>p_{\rm res})}{n_{\rm gas}} \frac{V_d - V_A}{V_A} \,.
\end{equation}
where $\Omega_{\rm ci}=e B_0/(m c)$ is the cyclotron frequency and $n_{\rm CR}(p>p_{\rm res})$ accounts only for particles with momentum larger than the resonant one.
In case of SNR shocks, Eq.~(\ref{eq:Gamma_res}) can be specialized as a function of the CR acceleration efficiency, $\xi_{\rm CR}$. For parameters typical of SNR shocks ($\xi_{\rm CR} \simeq 0.1$, $u_{\rm sh}\approx 5000$ km s$^{-1}$, $B_0 \approx 1 \mu$G, $v_A \approx 10$ km s$^{-1}$) and using a power low spectrum $f_{\rm CR}(p) \propto p^{-4}$, the growth time is 
\begin{equation} \label{eq:tau_res}
  \tau_{\rm growth}  =  \frac{1}{\Gamma_{\rm res}} 
  		\simeq \frac{2}{3\pi} \frac{\Lambda}{\xi_{CR}} 
		\frac{\gamma_{\rm res}}{\Omega_{\rm ci}} 
  		\left( \frac{c}{u_{\rm sh}} \right)^2  \left( \frac{V_A}{u_{\rm sh}} \right)
		\sim \gamma_{\rm res} \times {\cal O}(10^5 {\rm sec})  \,,
\end{equation}
where $\gamma_{\rm res}$ is the Lorentz factor of resonant particles and $\Lambda = \ln(p_{\max} / m c)$.
One can see that the instability grows rapidly. The maximum level is reached right ahead of the shock and can be estimated considering that waves can grow for a maximum time equal to the advection time, $t_{\rm adv} = D_1/u_{\rm sh}^2$. This condition gives
\begin{equation} \label{eq:Fsaturation1}
  {\cal F}_0(k)  =  \frac{\pi}{2} \,  \frac{\xi_{\rm CR}}{\Lambda} \, \frac{u_{\rm sh}}{V_A} \,.
\end{equation}
For the same values used above, Eq.(\ref{eq:Fsaturation1}) predict ${\cal F}_0 \gg 1$. Nevertheless, one has to keep in mind that this result has been obtained using the quasi-linear theory which assumes $\delta B/B_0 \ll 1$. When this condition is violated, as predicted by Eq.(\ref{eq:Fsaturation1}), the excited waves are no longer Alfv\'en waves and propagate with a speed larger than $V_A$.
If one makes the calculation properly, accounting also for the modification that CRs induce on the plasma dispersion relation \cite[see][section~4.2]{Blasi_rew:2013}, the saturation level is considerably reduced and the final power spectrum at the shock location turns out to be
\begin{equation} \label{eq:Fsaturation2}
  {\cal F}_0(k)  =  \left( \frac{\pi}{6} \,
  		                   \frac{\xi_{\rm CR}}{\Lambda} \, 
		                   \frac{c}{u_{\rm sh}}    \right)^{1/2} \,.
\end{equation}
Using the the usual canonical values one finds ${\cal F} \lesssim 1$, hence the effect of efficient CR acceleration is such as to reduce the growth of the waves and limit the value of the self generated magnetic field to the same order of magnitude as the pre-existing large-scale magnetic field. 

At this point one may wonder how is it possible to reach $\delta B \gg B_0$ required to explain CR up to the knee. 
At the moment of writing, the answer to this question remains open. What is known is that CRs can excite other kinds of instabilities, beyond the resonant one. Among them the most promising in terms of producing strong amplification is the so called {\it non-resonant Bell instability}  \citep{Bell:2004,Bell:2005}. This instability results from the $\bf{j} \times \bf{B}_0$ force that the current due  to escaping particles produces onto the plasma (see Fig.~\ref{fig:Bell-instability}). The non resonant instability grows very rapidly for high Mach number shocks. However, the scales that get excited are very small compared with the gyration radii of accelerated particles. Hence, it is not clear if the highest energy particles can be efficiently scattered. 
Indeed, hybrid simulations seem to confirm that the non-resonant Bell instability grows much faster than the resonant one for Mach number $\gtrsim 30$ and produces ${\cal F}\gg1$ \citep{Reville-Bell:2012,Caprioli-Spitkovsky:2014}. The same simulations also show that the instability produces a complex filamentary structure \citep{Reville-Bell:2013,Caprioli-Spitkovsky:2013} which could be able to scatter particles efficiently.
Unfortunately, even with the most advances numerical techniques it is, at present, difficult to simulate the whole dynamical range needed to describe the complex interplay between large and small scales. Therefore the question whether particles can reach the {\it knee} energy remains open.

\begin{figure}[t]
\includegraphics[width=0.5\linewidth]{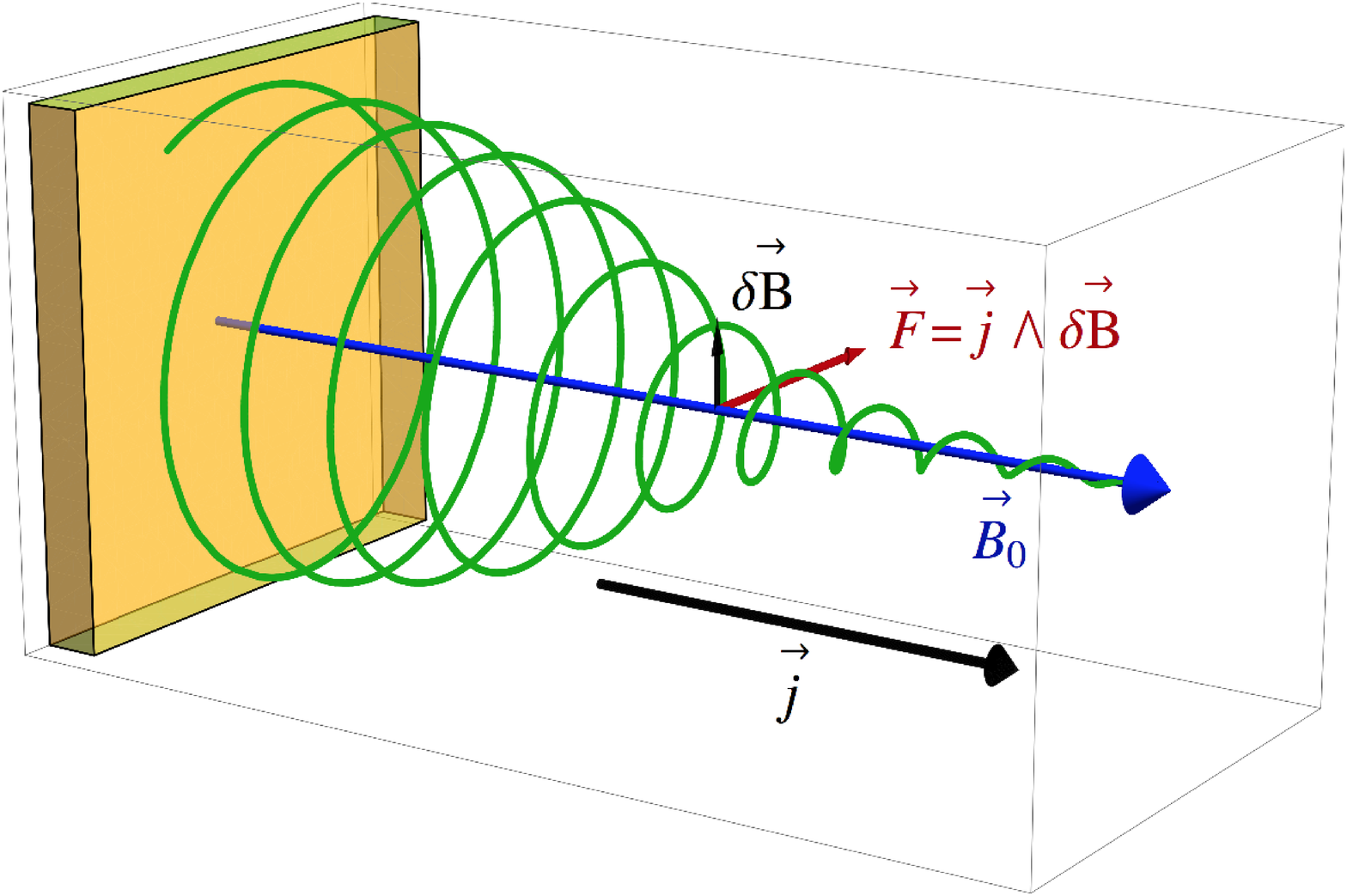}
\includegraphics[width=0.5\linewidth]{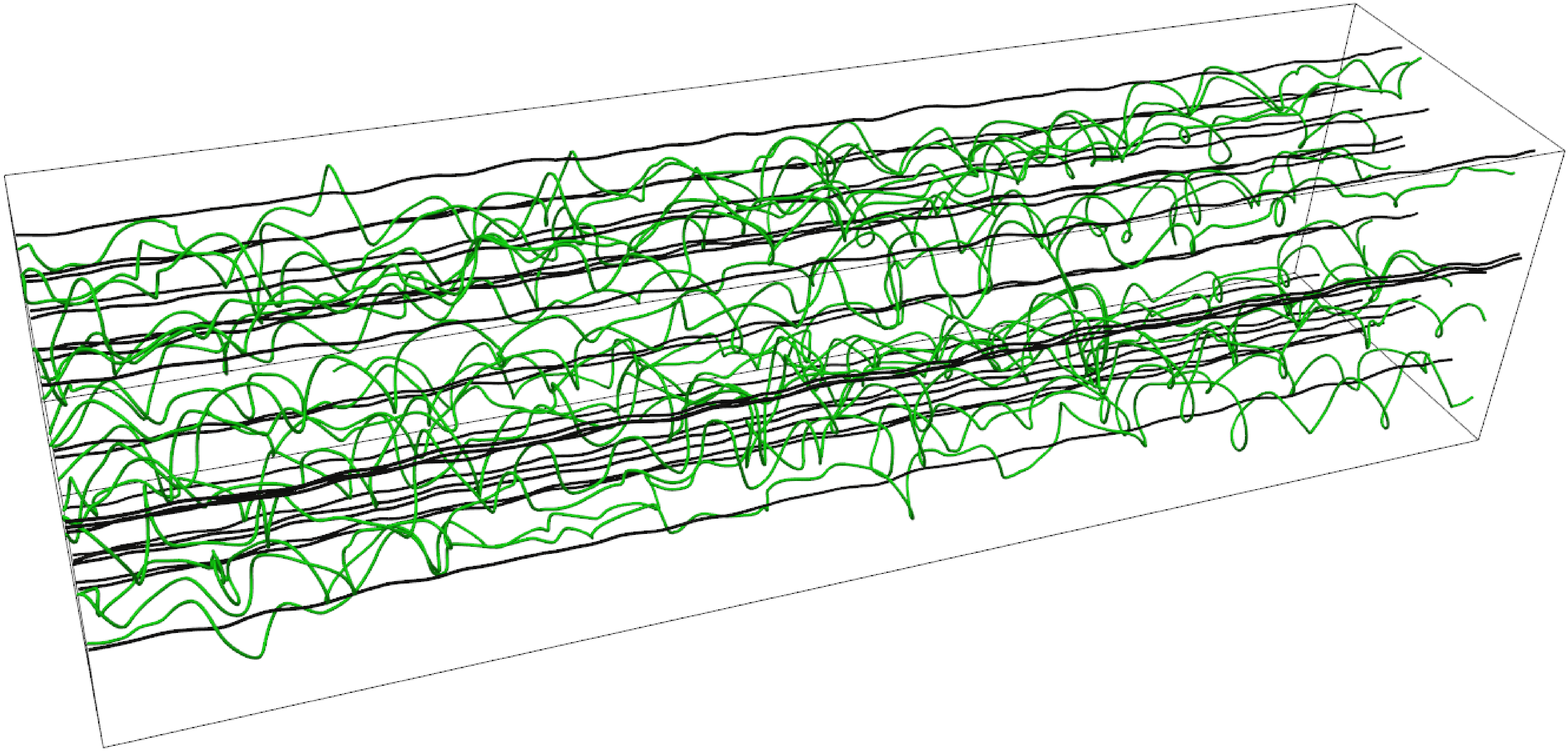}
\caption{{\bf Left}. Diagram showing how the non-resonant instability works: the current of CRs escaping from the shock region exerts a force ${\bf j} \times {\bf \delta B}$ onto the plasma, thereby stretching and amplifying the initial magnetic perturbations.
{\bf Right}. Result of non-resonant instability from a hybrid MHD simulation \citep{Reville-Bell:2013}. The approximately straight dark lines correspond to the current of escaping particles, while the green helical lines show the amplified magnetic field lines. Courtesy of Brian Reville.}
\label{fig:Bell-instability}
\end{figure}

\subsection{The dynamical reaction of the magnetic field}
\label{sec:morlino-MFreaction}
We anticipated that the magnetic field amplification upstream can resolve the problem of having very hard spectra in NLDSA.  
We saw that hard spectra result from a large compression factor which is in turn determined by the increased compressibility of the plasma. If the acceleration is absent or inefficient,  the compressibility is determined uniquely by the Mach number of the shock and the adiabatic index of the plasma (see Eq.~(\ref{eq:r})), while when acceleration is efficient the compressibility of the plasma increases essentially because the escaping particles are carrying away a non negligible fraction of the shock kinetic energy. 

The magnetic field can reduce the compression factor to values much closer to 4 (typically between 4 and 7) in two different ways. Firstly, if the magnetic field is amplified such that $\delta B \gg B_0$, the magnetic pressure may easily become larger than the upstream thermal pressure. The compression of the magnetic field component parallel to the shock surface modifies the shock jump conditions in such a way to reduce the compression factor. In other words the magnetic field makes the plasma ``stiffer'' \citep{Carpioli:2009}.

The second way to reduce the compression factor is through the damping of magnetic field \citep{McKenzie-Voelk:1982}, often called {\it turbulent heating}or {\it Alfv\'en heating}. Few mechanisms exist that can damp the magnetic field, the most relevant being the ion-neutral damping \citep{Kulsrud-Pearce:1969,Kulsrud-Cesarsky:1971} (which works only if a non negligible fraction of neutral hydrogen is present in the plasma) and the excitation of sound waves \citep{Skilling:1975b}. 
In all cases the final result of the damping is to convert a fraction of the magnetic energy into thermal energy, hence the plasma temperature increases and the Mach number of the shock is accordingly reduced.

It has been shown that the turbulent heating is less efficient than the magnetic compression in reducing the shock compression ratio. Moreover the former process has the strong inconvenience of reducing the strength of the magnetic field which is so precious to increase the maximum energy.
In passing we notice that acoustic waves can also be excited directly by CRs ({\it acoustic instability}) \citep{Drury-Falle:1986, Wagner-FH:2007} resulting in the plasma heating without requiring the damping of magnetic waves.

The mutual interplay between thermal plasma, magnetic field and accelerated particles described in this Section gives an idea of the complexity of NLDSA. A change in the compression factor due to CR pressure affects the spectrum of accelerated particles which in turn determines the level of magnetic field amplification, which also back reacts onto the shock structure.

\section{Escaping from the sources}
\label{sec:morlino-escaping}
In the test-particle picture of shock acceleration theory, accelerated particles are advected downstream of the shock and will be confined in the interior of the SNR until the shock disappears and the SNR merges into the ISM. At that point particles will be released in the ISM but they would have lost part their energy because of the adiabatic expansion of the remnant: hence the requirements in terms of maximum energy at the source would be even more severe than they already are. 
Therefore, effective escape from upstream, while the acceleration is still ongoing, is fundamental if high energy particles must be released in the ISM. 
The description of the particle escape from a SNR shock has not been completely understood yet, the reason being the uncertainties related to how particles reach the maximum energies \cite[a careful description of the numerous problems involved can be found in][]{Drury:2011}. Below we just describe the general framework.


Let us assume that the maximum momentum reached at the beginning of the Sedov-Taylor phase, $T_{\rm ST}$, is $p_{\max,0}$ and that then it drops with time as $p_{\max}(t) \propto (t/T_{\rm ST})^{-\beta}$, with $\beta > 0$. The energy in the escaping particles of momentum $p$ is
\begin{equation} \label{eq:f_esc}
  4 \pi f_{\rm esc}(p) pc p^2 dp = \xi_{\rm esc}(t) \frac{1}{2} \rho u_{\rm sh}^3 4 \pi R_{\rm sh}^2 dt
\end{equation}
where $\xi_{\rm esc}(t)$ is the fraction of the income flux, $\frac{1}{2} \rho u_{\rm sh}^3 4 \pi R_{\rm sh}^2$, that is converted into escaping flux.
If the expansion occurs in a homogeneous medium with $R_{\rm sh} \propto t^{\alpha}$ and $V_{\rm sh} \propto t^{\alpha-1}$, therefore, since $dt/dp \propto t/p$, from Eq.~(\ref{eq:f_esc}) we have:
\begin{equation} \label{eq:N_esc}
  f_{\rm esc}(p) \propto p^{-4} t^{5\alpha-2} \xi_{\rm esc}(t).
\end{equation}
It follows that in the Sedov-Taylor phase, where $\alpha=2/5$, the spectrum released in the ISM is $f_{\rm esc}(p) \propto p^{-4}$ if $\xi_{\rm esc}$ keeps constant with time. It is worth stressing that this $p^{-4}$ has nothing to do with the standard result of the DSA in the test-particle regime. Neither does it depend on the detailed evolution in time of the maximum momentum. It solely depends on having assumed that particles escape the SNR during the adiabatic phase.
Notice also that in realistic calculations of the escape $\xi_{\rm esc}$ usually decreases with time, leading to a spectrum of escaping particles which is even harder than $p^{-4}$. On the other hand, the total spectrum of particles injected into the ISM by an individual SNR is the sum of the escape flux and the flux of particles released after the shock dissipates.
This simple picture does not change qualitatively once the nonlinear effects of particle acceleration are included \citep{Caprioli-AB:2010}.

\section{The journey to the Earth}
\label{sec:morlino-propagation}

\begin{figure}[t]
\begin{center}
\includegraphics[width=0.8\linewidth]{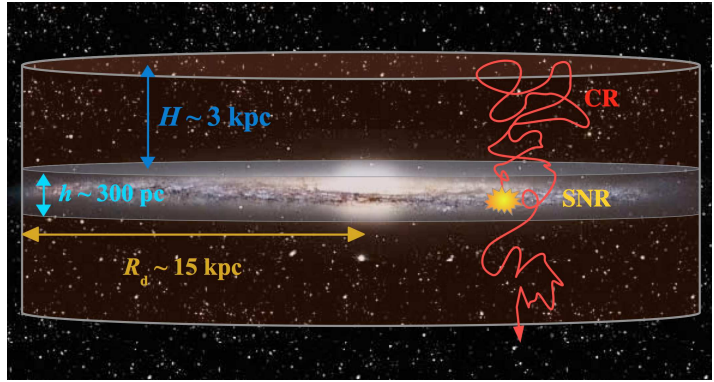}
\end{center}
\caption{Schematic representation of the {\it leaky box} model: CRs are produced by sources in the Galactic disk and diffuse in the magnetic halo above and below the disk, before escaping in the inter-galactic medium.}
\label{fig:halo}
\end{figure}

After leaving the sources, CRs start their journey through the Galaxy. When they arrive to the Earth, their incoming direction is nearly isotropic and does not mirror the distribution of matter in the Galaxy. This means that CRs diffuse toward us, loosing any information about the sources' location. The diffusion process that confines CRs in the Galaxy is believed to be due to the scattering by the irregularities in the Galactic magnetic field (the same described in Section~\ref{sec:morlino-Diffusion}), a process which depends on the particle's energy and charge $Z$. Therefore the CR spectrum at Earth results from the combination of injection and propagation. 

The basic expectation for how the spectrum at Earth relates to that injected by the sources is easily obtained in the so-called leaky box model, sketched in Fig.~\ref{fig:halo}. In this model the Galaxy is described as a cylinder of radius $R_d \approx 15$ kpc and thickness $h\approx 300$ pc, while a magnetized halo extents above and below the disk. The height of the magnetized Galactic halo is estimated from radio synchrotron emission to be $H\approx 3-4$ kpc. 
CRs are confined within this cylinder for a time $\tau_{\rm esc} \approx H^2 / D(E)$ with $D(E)$ the diffusion coefficient in the Galaxy. Let us write the latter as $D(E) = D_0 E^{\delta}$. If CR sources inject a spectrum $N_s(E) \propto E^{-\gamma_{\rm inj}}$ the spectrum of primary CRs at Earth will be:
\begin{equation} \label{eq:N_obs}
  N(E) \approx \frac{N_s(E) R_{\rm SN}}{2 \pi R_d^2 H} \, \tau_{\rm esc} \propto E^{-\gamma_{\rm inj} - \delta} \,,
\end{equation}
where $R_{\rm SN}$ is the rate of supernova explosion. Therefore what we measure at Earth only provides us with the sum of $\gamma_{\rm inj}$ and $\delta$. On the other hand, during their propagation in the Galaxy CRs undergo  spallation processes producing secondary elements: some of them, like boron, mostly result from these interactions. The spectrum of secondaries will be given by:
\begin{equation} \label{eq:N_sec}
N_{\rm sec}(E) \approx N(E) \, R_{\rm spall} \, \tau_{\rm esc} \propto  E^{-\gamma_{\rm inj}- 2 \delta}
\end{equation}
where $R_{\rm spall}$ is the rate of spallation reactions. It is clear then that the ratio between the flux of secondaries and primaries at a given energy $N_{\rm sec}(E)/N(E) \propto E^{-\delta}$ can provide us with a direct probe on the energy dependence of the Galactic diffusion coefficient and hence allow us to infer the spectrum injected by the sources.

A compilation of available measurements of the Boron-to-Carbon ratio is shown in Fig.~\ref{fig:BC} as a function of energy per nucleon. It is immediately apparent from the figure that the error bars on the high energy data points are rather large, and leave a considerable uncertainty on the energy dependence of the diffusion coefficient, being compatible with anything in the interval $0.3 < \delta < 0.6$. As a consequence, the slope of the CR spectrum at injection is also uncertain in the interval $2.1 < \gamma_{\rm inj} < 2.4$.
Notice that the ratio $N_{\rm sec}(E)/N(E)$ also provides the absolute value of the escaping time, because the spallation rates are known, giving $\tau_{\rm esc} \sim 5 \times10^6$ yr at 1 GeV.

\begin{figure}[t]
\begin{center}
\includegraphics[width=0.9\linewidth]{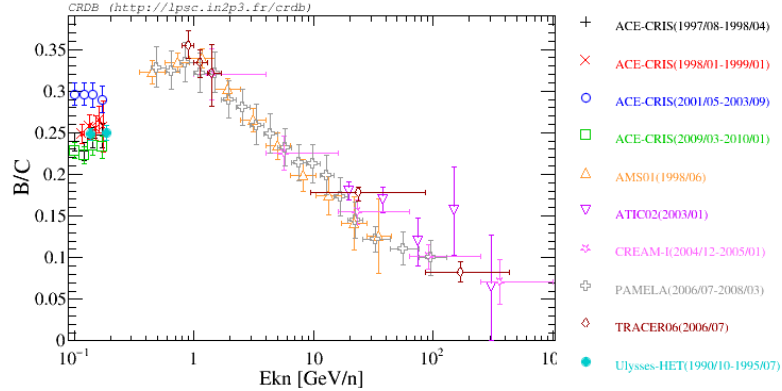}
\end{center}
\caption{Boron over Carbon ratio as a function of energy per nucleon taken from several experiments. Data have been extracted from the Cosmic Ray Database \cite[http://lpsc.in2p3.fr/crdb/,][]{Maurin:2014}.}
\label{fig:BC}
\end{figure}

This rather simple picture of CR transport through the Galaxy is complicated by other phenomena. The most relevant one is probably the possible presence of a large scale Galactic wind which can advect particle far away from the galactic plane \citep{Breitschwerdt-MV:1991,Zirakashvili-BPV:1996,Recchia-BM:2016}.

\section{Observational evidence}
\label{sec:morlino-obs}
In this Section we list the most relevant observations that support the idea that SNRs are indeed the main factories of Galactic CRs. To be clear, there are no doubts that SNR shocks are able to accelerate particles as will be clear from below. 
The question is rather to understand whether SNRs (and specifically which type of SNR and in which evolutionary phase) can explain all, or almost all, the observed CR flux, including particles up to $\approx 10^{17}$ eV. In this sense the evidence we have gathered until now is only circumstantial.
\newline
\newline
{\bf (1) Synchrotron emission from electrons.}
\newline
Multi-wavelength observations of young SNRs from radio to X-rays clearly show a dominant component of non-thermal emission which can only be explained as due to  synchrotron radiation emitted by highly relativistic electrons ($E\gtrsim1$ TeV). Most  SNRs have a radio spectral index $-0.6\lesssim \alpha_{\rm syn} \lesssim -0.4$, implying that electron energy spectrum resembles a power law $\propto E^{-s}$ with a spectral index $1.8 \lesssim s \lesssim 2.2$ with an average value of 2.0 \citep{Green:2014,Reynoso-Walsh:2015}.  The SNR morphology also shows that the highest energy emission occurs predominantly at the forward shock (see Fig.\ref{fig:SNR-Xrays}). Because DSA does not distinguish between leptons and hadrons, being dependent only on the particle's rigidity, there is no obvious reason to think that only electrons are accelerated. 
\newline
\newline
{\bf (2) Gamma radiation.} 
\newline
Accelerated hadrons can be detected through the decay of neutral pion produced when CR protons (or heavier nuclei) collide with the surrounding gas, i.e. $p_{\rm CR} p_{\rm gas} \rightarrow \pi^0 \rightarrow \gamma \gamma$. Unfortunately such emission occurs in the same energy range produced by electrons through the inverse Compton scattering of background photons. Analyzing the multi-wavelength spectrum, several studies have shown that the gamma-ray emission from some SNRs is better accounted for by hadronic models.
Furthermore gamma-ray emission has been detected also from a few molecular clouds close to SNRs, a fact interpreted as due to CRs escaping from the remnants and colliding with the high density region of molecular clouds, resulting in a strong production of $\pi^0 \rightarrow \gamma \gamma$. Specifically in two cases, SNRs IC443 and W44, the gamma-ray emission presents a low energy cut-off around $\approx 280$ MeV, a feature coinciding with the energy threshold of the $\pi^0$ decay \citep{Ackermann:2013}. 
\newline
\newline
{\bf (3) Signatures of an amplified magnetic field.}
\newline
In the last few years, the {\it Chandra} telescope has allowed us to measure the thickness of the X-ray emitting regions in SNRs (blue filaments shown in Fig.~\ref{fig:SNR-Xrays}), showing that in a number of remnants this is extremely compact, of order of 0.01 pc \cite[see][for recent reviews]{Vink:2012, Ballet:2006}. The simplest interpretation of these thin rims is in terms of synchrotron burn-off: the emission region is thin because electrons lose energy over a scale that is of order $\sqrt{D \tau_{\rm sync}}$, where $D$ is the diffusion coefficient and $\tau_{\rm sync}$ is their synchrotron lifetime. Assuming Bohm diffusion, this length turns out to be independent of the particle's energy and requires that the magnetic field responsible for both propagation and losses be in the 100 $\mu$G range.
Another observation that led to infer a large magnetic field is that of fast time-variability of the X-ray emission in SNR RX J1713.7-3946 \citep{Uchiyama-RXJ17:2007}. Again a field in the 100$\mu$G-1mG range was estimated, interpreting the variability time-scale as the time-scale for synchrotron losses of the emitting electrons.
Such high fields are strongly suggestive of efficient acceleration and of the development of related instabilities. However it should be mentioned that also alternative interpretations are possible \citep{Bykov-ER:2012,Schure-BDB:2012}. For example their origin might be associated to fluid instabilities that are totally unrelated to accelerated particles \citep{Giacalone-Jokipii:2007}. Therefore while the evidence for largely amplified fields seems very strong, it cannot be considered as a definite proof of efficient CR acceleration.
Nevertheless, it is worth mentioning that, at least in the case of SN1006, {\it Chandra} observation of the pre-shock region suggests that this amplification must be induced in the upstream \citep{Morlino-SN1006:2010}.
\newline
\begin{figure}[t]
\begin{center}
\includegraphics[width=0.8\linewidth]{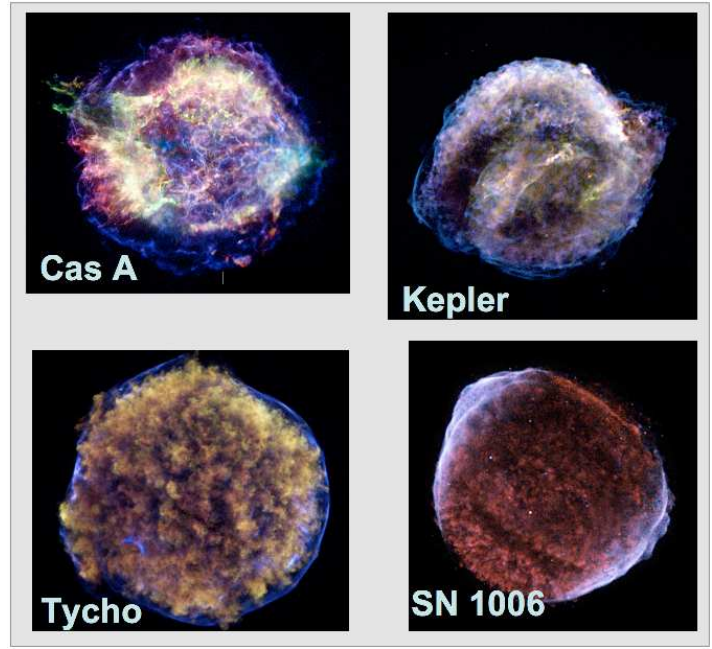}
\end{center}
\caption{A collection of X-ray images of young SNRs observed with the {\it Chandra} telescope. The blue color corresponds to the hard X-ray band (4-6 keV) where the emission is non-thermal. In this energy band thin filaments are clearly visible all around the remnants. They are interpreted as due to synchrotron emission of high energy electrons ($\sim 10$ TeV) in a strong amplified magnetic field ($B\sim 100-500 \mu$G). Image credit ``NASA/CXC''. For each single image:
Rutgers/G.~Cassam-Chena\"i, J.~Hughes et al. (SN 1006), SAO/D.~Patnaude et al. (Cas A), NCSU/S.~Reynolds et al. (Kepler) and CXC/Rutgers/J.~Warren \& J.~Hughes et al. (Tycho).}
\label{fig:SNR-Xrays}
\end{figure}
\newline
{\bf (4) Compression ratios.}
\newline
We have indeed evidence in at least two young SNRs, Tycho and SN 1006 \citep{Warren:2005,Cassam-Chenai:2008} that the distance between the contact discontinuity and the forward shock is smaller that that predicted by the  Rankine-Hugoniot jump conditions, that leads to infer a compression ratio of order seven in both cases. This value of the compression ratio is in  agreement with the predictions of NLDSA for the case of a shock that is efficiently accelerating particles and in which, either efficient turbulent heating takes place in the precursor, or the magnetic field is amplified to levels that make its energy density comparable with that of the thermal plasma upstream (see Section~\ref{sec:morlino-MFreaction}).
\newline
\newline
{\bf (5) Optical lines from the shocks.}
\newline
The last evidence we want to comment concerns a recent development of the DSA for shocks propagating in a partially ionized plasma \citep{Morlino-Ha3:2013}. In such conditions neutral hydrogen atoms can produce Balmer emission \index{Balmer lines} with a peculiar shape formed by two distinct lines, one narrow and one broad \citep{Chevalier-Raymond:1978,Chevalier-KR:1980} \cite[see also the review by][]{Heng:2010}.
SNR shocks are collisionless and when they propagate in a partially ionized medium, only ions are heated up and slowed down, while neutral atoms are unaffected to first approximation. However, when a velocity difference is established between ions and neutrals in the downstream of the shock,  the processes of  charge exchange and  ionization are  activated and these explain  the  existence  of  two  distinct  lines:   the  narrow line  is  emitted  by  direct  excitation  of  neutral  hydrogen after entering the shock front while the broad line results from the excitation of hot hydrogen population produced by charge-exchange of cold hydrogen with hot shocked protons. As a consequence, narrow and broad lines can directly probe the temperature upstream and  downstream of the shock, respectively.

Now, when the particle acceleration is efficient and a relevant fraction of kinetic energy is converted into relativistic particles, there is a smaller energy reservoir to heat the gas. Hence the downstream temperature turns out to be smaller than the case without acceleration (see how the thermal peak in Fig.~\ref{fig:modified-spectrum} moves towards lower energies as the shock efficiency increases). As a consequence the expected width of the broad Blamer line is smaller when efficient acceleration takes place.
It is very intriguing that such reduction of the broad Balmer line width has been inferred in at least two SNRs, namely RCW 86 and SN 0509-67.5 \citep{Morlino-Balmer:2014}. On the other hand, as shown in Section~\ref{sec:morlino-MFreaction}, the existence of a CR precursor could be responsible for a temperature increase of the upstream plasma resulting in a larger width of the narrow Balmer lines. Indeed, such anomalously larger width has been detected in several SNRs \citep{Sollerman-GLS:2003}.

\section{Conclusions}
\label{sec:morlino-conclusions}

This Chapter provides an overview of the basic physical ingredient behind the idea that SNRs are the main contributors to the Galactic CRs. 
The problem of the origin of CRs is a complex one: what we observe at the Earth results from the convolution of acceleration inside sources, escape from the sources and propagation in the Galaxy. Each one of these stages consists of a complex and often non-linear combination of pieces of physics.
A connection between SNRs and CRs was already proposed in the '30s on the basis of a pure energetic argument. Since then a complex theory has been developed where the particles are energized thorough a stochastic mechanism taking place at the SNR shocks. We have shown that the back reaction of accelerated particles onto the shock dynamics is the essential ingredient that allows particles to reach very high energies (probably up to $\approx10^{15}$ eV). 

From the observational point of view, there is enough circumstantial evidence suggesting that SNRs accelerate the bulk of Galactic CRs. This evidence is mainly based on the following pieces of observation: 
1) X-ray measurements show that SNRs accelerate electrons up to at least tens of TeV; 
2) gamma-ray measurements strongly suggest that SNRs accelerate protons up to at least $\approx 100$ TeV;
3) X-ray spectrum and morphology  show that magnetic field amplification is taking place at shocks of young SNRs, with field strength of order few hundred $\mu$.G This phenomenon is most easily explained if accelerated particles induce the amplification of the fields through the excitation of plasma instabilities.
4) In selected SNRs there is evidence for anomalous width of the Balmer lines, that can be interpreted as the result of efficient CR acceleration at SNR shocks.

A deeper look into the physics of particle acceleration will be possible with the upcoming new generation of gamma ray telescopes, most notably the Cherenkov Telescope Array (CTA). The increased sensitivity of CTA is likely to lead to the discovery of a considerable number of other SNRs that are in the process of accelerating CRs in our Galaxy. The high angular resolution will allow us to measure the spectrum of gamma ray emission from different regions of the same SNR so as to achieve a better description of the dependence of the acceleration process upon the environment in which acceleration takes place.

We conclude noticing that the CR physics should not be perceived as an isolated field of study, but has strong connection with other parts of Astrophysics. In fact CRs are an essential ingredient of the Interstellar Medium, their energy density being $\sim 1$ eV cm$^{-3}$, comparable with the energy density of other components (thermal gas, magnetic field and turbulent motion). This simple fact suggests that CRs can play a relevant role in many Galactic processes including the long term evolution of the Galaxy. In particular they are the only agent that can penetrate deep inside molecular clouds determining the cloud's ionization level and its chemical evolution, hence, directly affecting the initial condition of the star formation process. CRs can also be responsible for the generation of a Galactic wind which subtract gas from the Galactic plane, lowering the total star formation rate and polluting the inter-galactic medium with high metallicity gas. All these aspects represents open fields which promise interesting discoveries in the near future.


%
\begin{acknowledgement}
The author is grateful to Pasquale Blasi and Elena Amato for the long term collaboration on this subject and to Sarah Recchia and Marta D'Angelo for reading the manuscript.
\end{acknowledgement}

\bibliography{Morlino_References}

\end{document}